\documentclass{PoS}

\def\G{\Gamma}
\newcommand{\bea}{\begin{eqnarray}}
\newcommand{\eea}{\end{eqnarray}}

\title{Connecting amplitudes in different gauges beyond perturbation theory: a
canonical flow approach}

\ShortTitle{Yang-Mills amplitudes in different gauges in a canonical flow approach}

\author{\speaker{Andrea Quadri}\\
        Physics Dept., Universit\`a di Milano and INFN, Sez. di Milano\\
        via Celoria 16, I-20133 Milan, Italy\\
        E-mail: \email{andrea.quadri@mi.infn.it}}

\abstract{Physical quantities in gauge theories have to be gauge-independent. However their evaluation
can be greatly simplified by working in particular gauges. QCD provides several examples of this feature:
for instance, evolution equations in the Color Glass Condensate picture are usually derived in the Light-Cone gauge.
A more striking example is given by massive solutions of appropriate truncations to the QCD Schwinger-Dyson equations, that have been shown to exist in the Landau gauge, confirming lattice simulations carried out in the same gauge. 
Since physical quantities have to be gauge invariant, it is important to establish an approach allowing the comparison of computations carried out in different gauges even beyond perturbation theory. We show that the dependence on the gauge parameter $\alpha=0$ in Yang-Mills theories is controlled by a canonical flow that explicitly solves the Nielsen identities of the model. Green's functions in the $\alpha=0$ gauge are given by amplitudes evaluated in the theory at $\alpha=0$ (e.g., in the example of Lorentz-covariant gauges, in terms of Landau gauge amplitudes) plus some contributions induced by the $\alpha=0$-dependence of the generating functional of the canonical flow. Explicit formulas are presented and an application of the formalism to the gluon propagator is discussed.}

\FullConference{The European Physical Society Conference on High Energy Physics\\
		22--29 July 2015\\
		Vienna, Austria}

\begin{document}

\section{Introduction}

Gauge dependence of Green's functions in Yang-Mills theory is well-studied in the literature.
In the seminal papers~\cite{Nielsen:1975fs,Piguet:1984js} an algebraic approach to this problem has been
proposed, based on the so-called (generalised) Nielsen identities, i.e. a set of differential equations for the vertex functional $\G$
controlling the dependence on the gauge parameter $\alpha$. An application to the Standard Model has 
been given in~\cite{Gambino:1999ai}.

Formally the Nielsen identities can be obtained by extending the action of the BRST differential $s$ of the theory
to the gauge parameter $\alpha$ in such a way that it forms a BRST doublet~\cite{Barnich:2000zw,Quadri:2002nh} with a classical anti-commuting source $\theta$ (i.e. $s\alpha = \theta, s\theta =0 $).
This gives rise to an extended Slavnov-Taylor (ST) identity valid for the effective action $\G$.
The Nielsen identity is then easily recovered by taking a derivative w.r.t. $\theta$ of the ST identity.
Being based on symmetry arguments only, the ST identity is supposed to be valid even beyond perturbation theory.

This motivates the attempt to provide a general, purely algebraic method to obtain a solution of the ST identity order by order in $\alpha$, under the assumption of analyticity in the gauge parameter.

QCD provides  several examples where such a method could prove particularly useful.
The existence of massive solutions of appropriate truncations to the QCD Schwinger-Dyson equations~\cite{Aguilar:2008xm,Binosi:2009qm},
that has been established in the Landau gauge, confirming lattice simulations
again carried out in the Landau gauge, both in SU(2)~\cite{Cucchieri:2007md} and in
SU(3)~\cite{Bogolubsky:2009dc}.
Moreover, the study of the Kugo-Ojima function is also usually formulated
in the Landau gauge~\cite{Aguilar:2009pp}.
It is therefore desirable to have at disposal a method, to be applied in the non-perturbative regime, 
in order to allow the comparison of computations carried out in different gauges (see Ref.~\cite{Aguilar:2015nqa}
for the continuum and Ref.~\cite{Bicudo:2015rma} for lattice calculations connecting different linear gauges).

The formal treatment of the gauge parameter as a component of a BRST doublet establishes a close
analogy to the algebraic renormalization~\cite{Piguet:1995er} approach to the background field method.
There the background field $\hat A_\mu$ is paired with a classical anti-commuting source $\Omega_\mu$ under the BRST differential and an extended ST identity is obtained~\cite{Grassi:1995wr}-\cite{Ferrari:2000yp}.

Recently several advances have been made in order to provide an explicit solution of the  vertex functional
fulfilling such an extended ST identity order by order in the background field $\hat A_\mu$. It has indeed been proven~\cite{Binosi:2011ar}-\cite{Binosi:2012st} that the extended ST identity, reformulated in the context of the 
so-called Batalin-Vilkovisky (BV) formalism~\cite{Gomis:1994he}, induces a canonical (with respect to the BV bracket) transformation
that fixes uniquely the dependence of the vertex functional on the background $\hat A_\mu$.

The generating functional of such a canonical transformation is $\Psi_\mu = \frac{\delta \G}{\delta \Omega_\mu}$.
This functional depends on the background field $\hat A_\mu$ in a non-trivial way and this prevents to reconstruct the
full dependence of the vertex functional on the background by simple exponentiation. This in turn requires to make 
recourse to a certain Lie series~\cite{Binosi:2012st} in order to obtain the explicit coefficients of $\G$ order by order in powers of the background connection.  

This newly developed technique has allowed to obtain several results in Yang-Mills theories.
In the Color Glass Condensate picture, gauge-invariance of the evolution
equations has been proven to hold~\cite{Binosi:2014xua} as a consequence of the extended
ST identity, arising from the BRST symmetry of QCD in the presence
of the classical fast gluon backgrounds. Even though computations are usually
 carried out in the Landau gauge for the semi-fast gluons~\cite{Iancu:2000hn}-\cite{Hatta:2005rn}, any gauge choice for the
semi-fast modes can in fact be adopted.

Moreover, on the formal side, by exploiting the geometrical interpretation of the background field method
in the BV framework, it has been recognized~\cite{Binosi:2013cea} that 
the well-known antiBRST symmetry~\cite{Curci:1976bt}-\cite{Baulieu:1981sb}, which was early recognized in the literature but whose interpretation has remained rather mysterious, is indeed nothing but 
an equivalent reformulation of the background field method quantization of Yang-Mills theory.

In the present paper we summarize some recent progress~\cite{Quadri:2014jha} in the extension of the above tools to the 
study of the gauge dependence of Yang-Mills theory.
We discuss first the canonical flow
induced by the extended ST identity in the space of gauge parameters. 
The solution to this canonical flow is derived by means of the Lie series connecting 
the 1-PI amplitudes at $\alpha = 0$ (e.g., for Lorentz-covariant gauges, Landau gauge
Green's functions) with those at $\alpha \neq 0$. The Lie series contains some contributions
induced by the $\alpha$-dependence of the generating functional of the canonical flow.

We illustrate these results on the case of the gluon propagator in the Landau and in the Lorenz-covariant gauge.


\section{Classical Yang-Mills Action}

Let us consider pure SU(N) Yang-Mills theory with classical action
\bea
S=-\frac{1}{4g^2} \int d^4x \,  G_{a\mu\nu}^2 \, .
\label{cl.1}
\eea
The field strength is
\bea
G_{a\mu\nu} = \partial_\mu A_{a\nu} - \partial_\nu A_{a\mu} + 
f_{abc} A_{b\mu} A_{c\nu}
\label{cl.2}
\eea
and $f_{abc}$ the SU(N) structure constants.

The usual quantization procedure based on the BRST symmetry
requires the introduction in the tree-level vertex functional 
of a gauge-fixing function ${\cal F}_a$ through the
coupling with the Nakanishi-Lautrup multiplier field $b_a$:
\bea
S_{g.f.} = - \int d^4x \, b_a {\cal F}_a 
\label{cl.3}
\eea

We do not need to specify the exact form
of the gauge-fixing function ${\cal F}_a$. The only condition
is that it should allow the inversion of the tree-level 2-point functions
in the $A_{a\mu}-b_b$ sector.
${\cal F}_a$ might also depend on some parameters $\lambda_i$.
For instance one could interpolate  between the Lorentz-covariant gauge ($\lambda =0$) and 
the Coulomb gauge ($\lambda=1$) by choosing
\bea
{\cal F}_a = (1 - \lambda) \partial^\mu A_{a\mu} + \lambda~ \partial^i A_i
\label{cl.4}
\eea
Another example is the Slavnov-Frolov regularization of the Light Cone gauge
\bea
{\cal F}_a = A_- + \lambda \partial_- A_-
\eea
where $A_- = A_0 - A_3$ and $\partial_- = \partial_0 - \partial_3$.
Green functions are evaluated at $\lambda \neq 0$ and then one takes
the limit $\lambda \rightarrow 0$.

The analysis also applies to the case of a $R_\alpha$-gauge, when one adds
to the action the BRST-invariant term $\int d^4x \, \frac{\alpha}{2} b_a^2$ and study the Green's functions in different Lorentz-covariant gauges.

\section{BRST Symmetry}
Gauge invariance lost after the gauge-fixing procedure
is promoted to full BRST symmetry by adding 
to the classical action
both the gauge-fixing and the ghost-dependent terms
\bea
S_{g.f.+ gh} 
= s \int d^4x \, \bar c_a \left ( \frac{\alpha}{2} b_a - {\cal F}_a \right ) 
= \int d^4x \, \left ( \frac{\alpha}{2} b_a^2 - b_a {\cal F}_a 
+ \bar c_a s {\cal F}_a \right ) \, .
\label{cl.gfANDgh}
\eea

On the gauge field $s$ equals the gauge transformation, upon replacement of the gauge parameters with the ghost fields $c_a$
\bea
s A_{a\mu} \equiv D_\mu c_a = \partial_\mu c_a + f_{abc} A_{b\mu} c_c \, ,
\label{brst.a}
\eea
where $D_\mu c_a$ is the covariant derivative of the ghost field.
Moreover by nilpotency of $s$ one has
\bea
s c_a = -\frac{1}{2} f_{abc} c_b c_c \, .
\label{brst.c}
\eea
The antighost $\bar c_a$ and the Nakanishi-Lautrup multiplier field $b_a$ form a BRST doublet, i.e.
\bea
s \bar c_a = b_a \, , ~~~~~ s b_a = 0 \, .
\eea
The parameter $\alpha$ reduces to the usual gauge parameter for Lorentz-covariant gauges when ${\cal F}_a = \partial A_a$. The BRST symmetry is extended on the gauge parameters 
by setting
\bea
s \lambda_i = \theta_i \, , \quad s \theta_i = 0 \, , 
\quad s \alpha = \theta \, \quad s \theta = 0 \, .
\label{brst.par}
\eea

\section{Slavnov-Taylor and Nielsen Identities}
         
The full tree-level vertex functional
\bea
\G^{(0)} = S + S_{g.f.+gh} + S_{a.f.}
\eea
where $S_{g.f.+gh}$ is the gauge-fixing and ghost part and $S_{a.f.}$ is the antifield-dependent sector, with the couplings of the BRST variations of the fields to the corresponding external sources, called antifields, obeys the following Slavnov-Taylor (ST) identity, as a consequence of BRST invariance:
\bea
{\tilde{\cal S}}(\G^{(0)})  = 
\sum_i \theta_i \frac{\partial \G^{(0)}}{\partial \lambda_i} 
+ \theta \frac{\partial \G^{(0)}}{\partial \alpha} 
+
\int d^4x \, \Big (
\frac{\delta \G^{(0)}}{\delta A^*_{a\mu}} 
\frac{\delta \G^{(0)}}{\delta A_{a\mu}} 
-
\frac{\delta \G^{(0)}}{\delta c^*_{a}} 
\frac{\delta \G^{(0)}}{\delta c_{a}}
+
b_a \frac{\delta \G^{(0)}}{\delta \bar c_a} \Big )
 = 0 \, .
\label{ext.sti}
\eea
For non-anomalous theories this equation holds for the full vertex functional~$\G$.

By taking a derivative w.r.t. $\theta$ and then setting $\theta, \theta_i$ equal to zero one obtains the following Nielsen identity
\bea
\left . 
 \frac{\partial \G}{\partial \alpha}
\right |_{\theta=\theta_i=0}
 & = & -
\int d^4x \, \Big (
\frac{\delta^2 \G}{\partial \theta \delta A^*_{a\mu}} 
\frac{\delta \G}{\delta A_{a\mu}} 
-
\frac{\delta \G}{\delta A^*_{a\mu}} 
\frac{\delta^2 \G}{\partial \theta \delta A_{a\mu}} 
\nonumber \\
& & \left . -
\frac{\delta^2 \G}{\partial \theta \delta c^*_{a}} 
\frac{\delta \G}{\delta c_{a}}
-
\frac{\delta \G}{ \delta c^*_{a}} 
\frac{\delta^2 \G}{\partial \theta \delta c_{a}}
+
b_a \frac{\delta^2 \G}{\partial \theta \delta \bar c_a} \Big ) 
\right |_{\theta=\theta_i=0}
\, .
\label{nielsen.id}
\eea
A similar equation holds for the derivative of $\G$ w.r.t.
$\lambda_i$, once one takes a derivative of the extended ST
identity w.r.t. $\theta_i$.

\section{Canonical Flow}

In order to make it apparent the canonical flow solution 
of the Nielsen identity we rewrite the extended
ST identity within the Batalin-Vilkovisky (BV) formalism.
Hence one introduces an antifield $\bar c^*_a$ for the antighost
$\bar c_a$ as well as the antifield $b^*_a$ 
for the Nakanishi-Lautrup field $b_a$.
$\bar c^*_a$ is coupled to $b_a$ in the classical action, while
$b^*_a$ does not enter into $\G^{(0)}$ (since $s b_a = 0$).

The BV bracket is
\bea
&& 
\!\!\!\!\!\!\!\!\!\!\!\!\!\!\!\!\!\!\!\!\!\!\!
\{ X, Y \}  =  \int d^4x \, \sum_\phi \Bigg [
(-1)^{\epsilon_\phi (\epsilon_X+1)} \frac{\delta X}{\delta \phi} 
\frac{\delta Y}{\delta \phi^*} - (-1)^{\epsilon_{\phi^*} (\epsilon_X +1 )}
\frac{\delta X}{\delta \phi^*}\frac{\delta Y}{\delta \phi} \Bigg ] \, .
\label{bv.bracket}
\eea
The sum runs over the fields $\phi = (A_{a\mu}, c_a, \bar c_a, b_a)$
and the corresponding antifields $\phi^* = (A^*_{a\mu}, c^*_a, \bar c_a^*, b_a^*)$.

The extended ST identity can be written as
\bea
\tilde{\cal S}(\G) = 
\sum_i \theta_i \frac{\partial \G}{\partial \lambda_i} 
+ \theta \frac{\partial \G}{\partial \alpha} + \frac{1}{2} \{ \G, \G \} = 0
\, .
\label{st.bv}
\eea
By taking a derivative w.r.t. $\theta$ one finds
\bea
\left . \frac{\partial \G}{\partial \alpha} \right |_{\theta = \theta_i =0} 
= - \left . \{ \frac{\partial \G}{\partial \theta}, \G \} 
\right |_{\theta = \theta_i =0} \, .
\label{eq.alpha}
\eea
This equation shows that the derivative of the vertex functional
w.r.t. $\alpha$ is obtained by a canonical transformation 
(w.r.t. the BV bracket) induced by the 
generating functional $\Psi \equiv \frac{\partial \G}{\partial \theta}$.
Since the latter in general depends on $\alpha$, one cannot
solve Eq.(\ref{eq.alpha}) by simple exponentiation and one needs
to make recourse to a Lie series.

\section{The Gauge Lie Series}

We introduce the operator
\bea
\Delta_\Psi = \{ \cdot, \Psi \} + \frac{\partial}{\partial \alpha} \, .
\label{Delta}
\eea
Then the vertex functional $\G$ is given by the following Lie
series~\cite{Binosi:2012st}
\bea
\G = \sum_{n \geq 0} \frac{1}{n!} \alpha^n [\Delta_\Psi^n \G_0]_{\alpha = 0}
\label{Lie.series}
\eea
where $\G_0$ is the vertex functional at $\alpha=0$.
Notice that one must afterwards take the limit $\alpha \rightarrow 0$
(although the operator $\Delta_\Psi$ is applied on the functional
$\G_0$, which is $\alpha$-independent) since a residual $\alpha$-dependence
may arise (and in general indeed arises) from the differentiation w.r.t. $\alpha$ 
of the generating functional $\Psi$.

\section{The Gluon Propagator}

As an example, let us consider in the perturbative regime 
how one can derive the solution to the gauge evolution equation
for the transverse part of the 
gluon propagator. For that purpose 
we introduce the transverse and longitudinal form factors according to
\begin{eqnarray*}
\Delta_{A^a_\mu A^b_\nu} = -i \delta^{ab} \Big ( 
\Delta_T(p^2) T^{\mu\nu} + \Delta_L(p^2) L^{\mu\nu} \Big )  \, .
\label{gluon.1}
\end{eqnarray*}
The relevant quantity is $\Delta_T(p^2)$. 
The canonical flow equation is in this case
\bea
\partial_\alpha \G_{A_{b_1 \nu_1} A_{b_2 \nu_2}}
= - \int d^4x \,\Big [  \G_{\theta {\tilde A}^*_{a\mu} A_{b_1 \nu_1}} \G_{A_{b_2 \nu_2} A_{a\mu}}
+ \G_{\theta {\tilde A}^*_{a\mu} A_{b_2 \nu_2}} \G_{A_{b_1 \nu_1} A_{a\mu}} \Big ] \, .
\label{2pt}
\eea
Short-hand notation where lowstair letters denote functional differentiation w.r.t. that argument and
it is understood that in the end one sets all fields $\Phi$ 
and external sources $\Phi^*,\theta,\theta_i$ equal to zero. For instance
\bea
\G_{A_{b_1 \nu_1} A_{b_2 \nu_2}} \equiv \left. \frac{\delta^2 \G}{\delta A_{b_1 \nu_1} 
\delta A_{b_2 \nu_2}} \right |_{\Phi=\Phi^*=\theta=\theta_i=0} \, .
\label{1pi.fnct}
\eea
Let us introduce transverse and longitudinal form factors for 
the 1-PI functions involved, namely (in the Fourier space)
\bea
&& \G_{A_{b_1 \nu_1} A_{b_2 \nu_2}} = \delta_{b_1 b_2} \Big ( G^T T_{\mu\nu} + G^L L_{\mu\nu} \Big ) \, , \nonumber \\
&& \G_{\theta {\tilde A}^*_{a\mu} A_{b \nu}} = \delta_{a b} \Big ( R^T T_{\mu \nu}
+ R^L L_{\mu\nu} \Big ) \, .
\label{form.fact}
\eea
Then by applying the transverse projector to Eq.(\ref{2pt}) one gets
\bea
\frac{\partial G^T}{\partial \alpha} = - 2 R^T G^T \, .
\label{diff.eq}
\eea
Let us denote by $G^T_0$ the form factor in the Landau gauge.
Then by integrating Eq.(\ref{diff.eq}) one gets
\bea
G^T = \exp \Big ( - \int_0^\alpha 2 R^T ~ d\alpha' \Big ) G^T_0
\eea
and therefore for the transverse part of the gluon propagator
\bea
\Delta^T =\exp \Big ( \int_0^\alpha ~ 2 R^T ~ d\alpha' \Big ) \Delta ^T_0 \, .
\label{rel.prop}
\eea

On the other hand, by Eq.(\ref{rel.prop}) the following ratio 
\bea
r =  \exp \Big ( - \int_0^\alpha 2 R^T ~ d\alpha' \Big ) \frac{\Delta^T}{\Delta^T_0}
\label{ratio}
\eea
must be equal to one (and therefore gauge-independent).

\section{Conclusions}

The existence of a canonical flow in the space of gauge parameters
and the related solution in terms of a Lie series provide a way
to compare results in different gauges within an algebraic framework
that is bound to hold even beyond perturbation theory (as far as 
the ST identity is valid).

The dependence of the generating functional of the canonical flow
on the gauge parameter prevents to get the full solution
by a naive exponentiation. Such a solution can be expressed
through an appropriate Lie series, in close analogy to the
solution of the extended ST identity in the presence of
a background gauge connection.

Knowing such a Lie series eases the comparison
between computations carried out in different gauges. In the
simplest example of the 2-point gluon function, a closed formula
interpolating between the Landau and the Lorentz-covariant gauge
can be obtained, under the assumption that analyticity in the 
gauge parameter around $\alpha=0$ holds.

\end{document}